\documentclass[aps, 10pt, twocolumn,preprintnumbers,prd, amsmath,amssymb,a4page,notitlepage]{revtex4} 
\usepackage{graphicx}
\usepackage{bm}
\usepackage{amssymb}
\usepackage{epsfig}
\usepackage{color}
\usepackage{mathtools}
\usepackage[
colorlinks=true,
filecolor=black,
anchorcolor=blue,
linkcolor=blue,
citecolor=cyan, 
urlcolor=cyan,
linktocpage=true,
plainpages=false,
breaklinks=true,
pdfstartview=FitH
]{hyperref}

\newcommand{\nn}{\nonumber}

\newcommand{\be}{\begin{equation}}
\newcommand{\ee}{\end{equation}}
\newcommand{\bea}{\begin{eqnarray}}
\newcommand{\eea}{\end{eqnarray}}

\begin{document}

\title{Light Dark Matter Axion Detection with Static Electric Field}

\author{Yu Gao$^{1}$}
\author{Yongsheng Huang$^{2}$}
\author{Zhengwei Li$^{1}$}
\author{Manqi Ruan$^{1,3}$}
\author{Peng Sha$^1$}
\author{Meiyu Si$^{1,3}$}
\author{Qiaoli Yang$^{4}$}

\affiliation{
$^{1}$Institute of High Energy Physics, Chinese Academy of Sciences, Beijing 100049, China\\
$^{2}$School of Science, Shenzhen Campus of Sun Yat-sen University, Shenzhen 518107, China\\
$^{3}$University of Chinese Academy of Sciences, Beijing 100049, China\\
$^{4}$Physics Department and Siyuan Laboratory, Jinan University, Guangzhou 510632, China\\
}

\begin{abstract}
We explore the axionic dark matter search sensitivity with a narrow-band detection scheme aiming at the axion-photon conversion by the static electric field inside a cylindrical capacitor. An alternating magnetic field signal is induced by effective currents as the axion dark matter flows perpendicularly through the electric field. At low axion masses, like in a KKLT scenario, front-end narrow band filtering is provided by using LC resonance with a high $Q$ factor, which enhances the detectability of the tiny magnetic field signal and also leads to a thermal noise as the major background that can be reduced at cryogenic conditions. We demonstrate that high $g_{a\gamma}$ sensitivity can be achieved by using a strong electric field. The QCD axion theoretical parameter space can be reached with high $E\sim$ GVm$^{-1}$ field strength. Using the static electric field scheme essentially avoids exposing the sensitive superconducting pickup to an applied laboratory magnetic field.
\end{abstract}
\maketitle

\section{Introduction}
\label{sect:intro}

Astrophysical and cosmological observations indicate the existence of cold dark matter as 23\%~\cite{Ade:2015xua} of the Universe's total energy budget. Promising scenarios includes the dark matter in form of extra-low mass bosons~\cite{Turner:1983he}, and one highly motivated candidate is the axion~\cite{Peccei:1977hh, Weinberg:1977ma, Wilczek:1977pj, Peccei:1977ur, Kim:1979if}. Naturally relaxing the amount of Charge-Parity violation in the Standard Model's strong interaction~\cite{Peccei:2006as,Kim:2008hd} (a.k.a. the strong CP problem ~\cite{Crewther:1979pi}), the axion emerges from a natural global $U(1)_{\rm PQ}$ symmetry extension of the Standard Model (SM), which is spontaneously broken at high scale and later explicitly broken at the strong interaction's confinement scale by the QCD instanton potential~\cite{Callan:1977gz, Vafa:1984xg}, making the axion a pseudo-Nambu-Goldstone particle with a small but non-zero mass. Cosmologically, through the `misalignment' mechanism~\cite{Preskill:1982cy, Abbott:1982af, Dine:1982ah,Sikivie:1982qv,Ipser:1983mw,Sikivie:2006ni} axions are produced during the QCD phase transition and their collective evolution make up the dark matter density observed at the current time. Depending on the spontaneous $U(1)_{\rm PQ}$ breaking scale relative to that of inflation, typically cosmic axions account for the relic density in the classical mass window with $m_a\sim {\cal O}(10^{-5}-10^{-3})$~eV~\cite{Sikivie:2006ni} or  $m_a<{\cal O}(10^{-7})$~eV~\cite{Hertzberg:2008wr}, and a wider dark matter axion mass range is possible through other production mechanisms.

The axion acquires its characteristic $\frac{a}{f_a}G\tilde{G}$ coupling through a loop of $SU(3)_c$-charged fermions, such as in benchmark invisible axion scenarios like the KSVZ~\cite{Kim:1979if, Shifman:1979if} and the DFSZ~\cite{Zhitnitsky:1980tq, Dine:1981rt} models. Generally the axion can also couple to other SM gauge fields, in particular the QED photon, with the interaction Lagrangian term
\be
{\cal L}_{a\gamma\gamma}=-g_{a\gamma}a\vec E\cdot \vec B~~,
\ee
where $\vec{E},\vec{B}$ are the electric and magnetic field strengths, the axion-photon coupling is $g_{a\gamma}=c_{\gamma}\alpha_{\rm QED} / (\pi f_a)$, with $f_a$ as the $U(1)_{\rm PQ}$ symmetry's scale factor, $\alpha_{\rm QED}=1/137$ is the fine-structure constant and the ${\cal O}(1)$ coefficient $c_{\gamma}$ is given by the details of the particular UV model. This coupling also characterizes the generalized axion-like particles (ALPs), which can be created from dimension compactification in string theory, or other UV theories with an extra light bosonic mode that may not necessarily couple to the SM gluon field. General ALPs have relaxed $f_a-m_a$ correlation compared to that of the QCD axion, thus they have a much wider parameter space to serve as a dark matter candidate. In the following we will use `axion' to denote both the QCD axion and ALPs if not specified.

The axion-photon coupling $g_{a \gamma}$ allows the dark matter axion convert into photons inside a laboratory's electromagnetic fields, and is being actively searched for by cavity haloscopes including ADMX~\cite{ADMX:2021nhd}, HAYSTAC~\cite{Zhong:2018rsr}, CAPP~\cite{CAPP:2020utb}, ORGAN~\cite{McAllister:2018xgn}, CAST-RADES~\cite{CAST:2020rlf}, QUAX-$a\gamma$~\cite{Alesini:2020vny}, TASEH~\cite{Chang:2022dql}, DANCE~\cite{Oshima:2021irp}, non-cavity experiments such as ABRACADARA~\cite{Salemi:2019xgl}, high-frequency dielectric design like MADMAX~\cite{Li:2021mep}, BREAD~\cite{BREAD:2021tpx}, and many others. For recent reviews of axion search methods, see Ref.~\cite{Irastorza:2018dyq,Sikivie:2020zpn}.
So far, resonant cavity haloscopes achieved the highest sensitivity due to the a high-$Q$ front-end enhancement at both classical~\cite{Sikivie:1983ip} and quantum levels~\cite{Yang:2022uil}. However, matching the signal photon's wavelength in the low axion mass range $m_a\ll10^{-6}$ eV, motivated by string theory~\cite{Svrcek:2006yi}, GUT scale new physics~\cite{Barbieri:1981rs,Giudice:2012zp,Hall:2014vga,Ernst:2018bib, Gao:2019tqt}, etc., faces practical limitation of cryogenic equipment sizes. For a low axion mass the size of the resonant cavity could be too large, and multiple promising complementary methods are helpful, such as in ABRACADABRA~\cite{Salemi:2019xgl}, SHAFT~\cite{Gramolin:2020ict}, ADMX-SLIC~\cite{ADMXSLIC}, BASE~\cite{BASE}, DMRadio~\cite{Brouwer:2022bwo}, etc., each featuring different strengths and limitations.   

In the theoretical perspective, $m_a\lesssim 10^{-7}$~eV axion does have its unique attractions. It has been shown that very light axions have strong implications to inflation~\cite{Hertzberg:2008wr}. Models with a very light axion typically have a PQ symmetry breaking scale $f_a$ order of the GUT scale, which results in considerable tensions between the observed isocurvature fluctuations and high-scale inflation. A low energy scale inflation scenario will be necessary if very light dark matter axions are detected. The discovery of very light dark matter axions will also indicate the observed Universe as a highly atypical Hubble volume and possibly the KKLT scenario~\cite{Hertzberg:2008wr}. Thus, for low-mass axions, an alternative detection method has been proposed in Ref.~\cite{Sikivie:2013laa} to utilize the high $Q$-factor of a resonant LC circuit as a narrow-band pickup. As the dark matter to searched for the axion-induced signal with a circuit resonance enhanced probe~\cite{Sikivie:2013laa, Brouwer:2022bwo,Salemi:2019xgl}. However the very strong magnetic fields created by magnets or solenoids inherently possess small fluctuations which would be hard to separate from the tiny axion induced signals.

In this paper we propose a narrow-band axion dark matter experimental scheme based on electrified cylindrical capacitor. We investigate the prospective sensitivity to axion-photon coupling $g_{a\gamma}$ with a resonance circuit as the high $Q$ pickup of induced signals from axion's coupling to strong laboratory electric field. Compared to strong magnetic fields, although axion's conversion rate in electric field is suppressed by a sub-unity local dark matter flux velocity, however, using the $\vec{E}$ field~\cite{Gao:2020sjn} avoids the presence of a strong magnetic field in close proximity of the signal pickup. In addition, the signal strength with an electric field is modulated by the relative angle of the $\vec{E}$ field orientation and the direction of the local dark matter flux, which provides a potential test of the origin of a signal.

In Section~\ref{sect:setup}, we describe the experimental schematics with a meter-scale cylindrical capacitor, and calculate the signal strength of the DM axion-induced signal with a LCR resonance pickup. Section~\ref{sect:sensitivity} discusses the major noises and the prospective sensitivity in an axion mass range $m_a< 10^{-6}$ eV. We demonstrate that good axion-photon coupling sensitiivty can be achieved with a static electric field as the converting medium, then we conclude in Section~\ref{sect:discussion}.

\begin{figure}[t]
\includegraphics[scale=0.52]{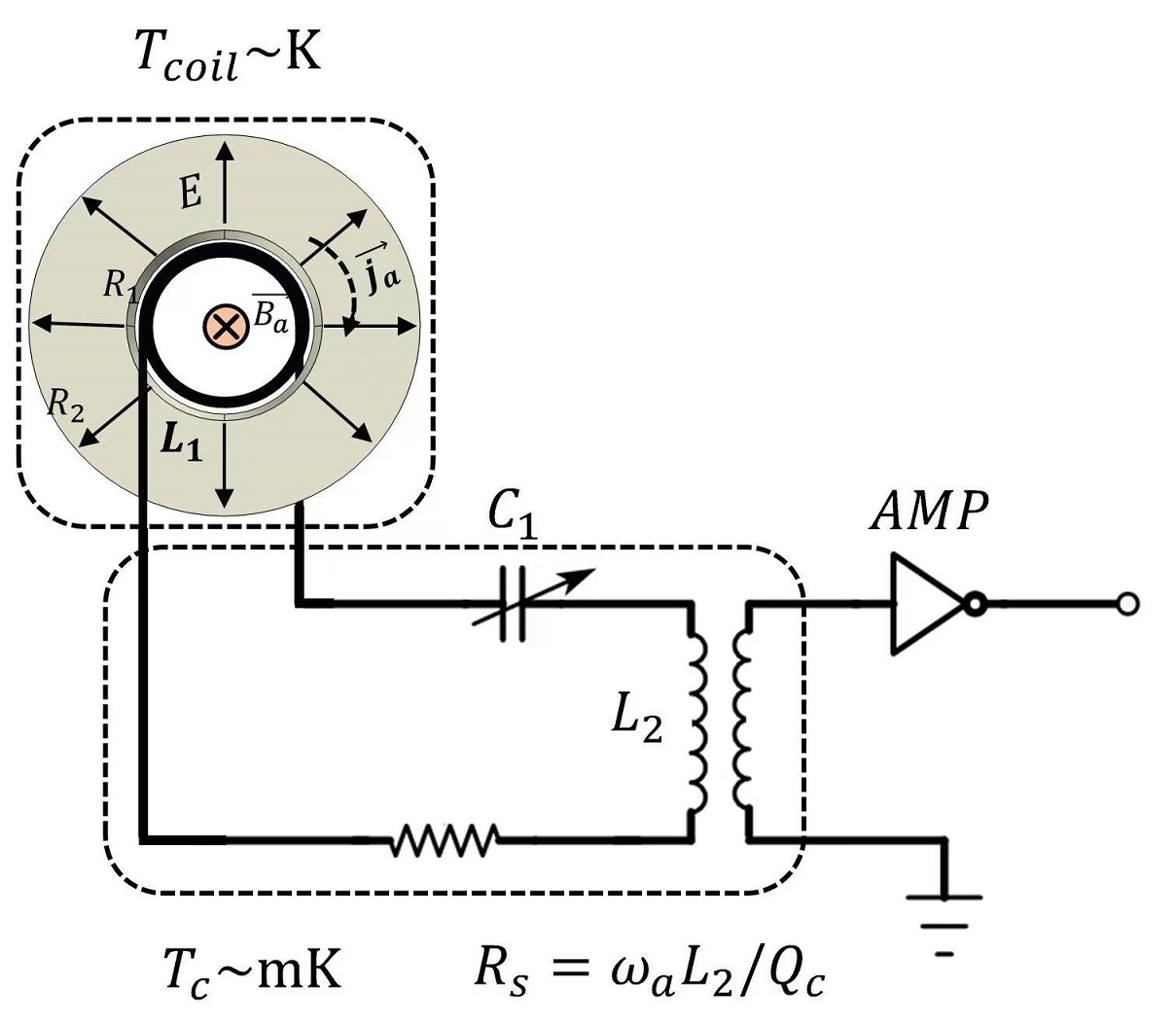}
\includegraphics[scale=0.27]{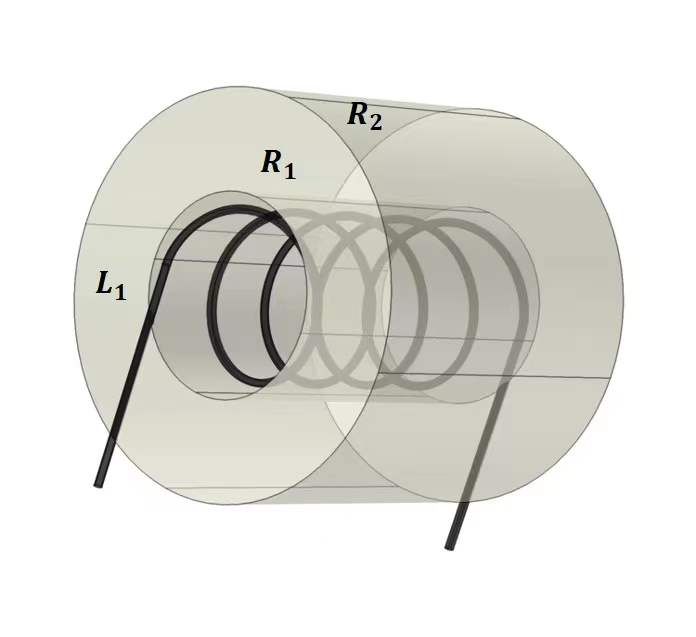}
\caption{Illustrative experiment setup. Upper: a superconductor coil $L_1$ is placed in the central region of a cylindrical capacitor, and is connected in a resonant LCR circuit as a narrow-band pickup for the axion-induced signal $\vec{B}_a$. The inner layer ($R_1$) of the capacitor should be non-conducting in order to let in the induced signal. $R_s$ denotes the total resistance. Lower: The pickup coil $L_1$ is loosely winded to reduce its self-inductance and allows the signal to penetrate into the central region. Using superconductor $L_1$ and $L_2$ can lower their resistance and reduce thermal noises.}
\label{fig:setup}
\end{figure}

\section{Cylindrical Capacitor Setup}
\label{sect:setup}

For terresital laboratories, the dark matter axion can be modeled as a plane wave
\be
a(x,t)\approx a_0{\rm cos}\left[m_a\vec v_a\cdot \vec x-\left(m_a+{m_a\over 2}v_a^2\right)t \right]~,
\ee
with a de-Broglie wavelength $\lambda_a= k_a^{-1}\sim (v_a m_a)^{-1}$. $v_a\sim 10^{-3}$ the Earth's relative velocity to the galaxy's dark matter halo, and the axion field magnitude is $a_0=\sqrt{2\rho_{\rm DM}}/m_a$ where the local DM halo density is typically $\rho_{\rm DM}=0.4$ GeV cm$^{-3}$. For $\lambda_a$ longer than the dimension of the experimental apparatus, this axion field can be considered as a coherent classical background field that oscillate at $\omega_a = (1+v_a^2/2)m_a$. The energy spread due to the halo's velocity dispersion is usually small, and it is often denoted by the axion field's quality factor as $\delta \omega_a /\omega_a \equiv Q_a^{-1}$. For conventional dark matter halo models, $Q_a \sim 10^{6}$, and any signals from this nearly monochromatic oscillation would be efficiently detected from a narrow-band probe with a matched quality factor.

When the axion is considered as a background field, the axion-modified Maxwell equations~\cite{Sikivie:1983ip} observe the addition of effective 4-current density $j^\mu=\{\vec\nabla a\cdot \vec{B} ,-g_{a\gamma}\vec E\times \vec \nabla a+g_{a\gamma}\dot{a}\vec B  \}$ that simplifies to an effective displacement current density $\vec{j}_a$ inside a static laboratory electric field. The modified Maxwell equations are
\bea
\vec\nabla\cdot \vec E&=&
\rho_e, \nonumber \\
\vec\nabla\times\vec B-{\partial \vec E \over \partial t}&=& \vec{j}_e + \vec{j}_a, \\
\vec\nabla\cdot\vec B&=&0,\nonumber\\
\vec\nabla\times \vec E&=&-{\partial \vec B\over \partial t}~, \nonumber
\eea
where $\vec{E}=\vec{E}_0+\vec{E}_a$ is dominated by the static field $\vec{E}_0$ and $\vec{j}_e =0$. $\vec{j}_a=-g_{a\gamma}a \vec E_0\times\vec{k}_a$ alternates at the same frequency as the axion field, and it induces a magnetic field signal $\vec{B}_a$. Since $\vec{j}_a$ is perpendicular to both $\vec{E}$ and $\vec{v}_a$, $\vec{j}_a$ depends by the angle between the local dark matter density flow density and the applied field $\vec{E}_0$. This will lead to periodic $\vec{j}_a$ modulations, for instances, by 24 hours due to the Earth's rotation, and by $2\pi(v_a \omega_a)^{-1}$ if $\lambda_a$ is significantly longer than the size of the $\vec{E}_0$-field region. Also note its amplitude $j_a= g_{a\gamma}\sqrt{2\rho_{\rm DM}}\cdot|\vec{v}_a\times \vec{E}_0|$ is insensitive to $m_a$ for a given dark matter energy density.

With a cylindrical capacitor, as illustrated in Fig.~\ref{fig:setup}, and assuming $\vec{a}$ is along the cylinder axis, $\vec{j}_a$ is circularly distributed in the radial $\vec{E}$-field region between the capacitor plates, forming an effective current loop and the induced magnetic field amplitude along the cylinder axis~\cite{Gao:2020sjn} is
\bea
B_a &=& g_{a\gamma}E_0 v_{\rm DM} c_R\sqrt{2\rho_{\rm DM}}R_1 \\
&\sim&2\times 10^{-10} {\rm T} \cdot \left( \frac{g_{a\gamma}}{\rm GeV^{-1}}\right)
\left(\frac{E_0}{10^7 {\rm V/m}} \right)
\left(\frac{R_1}{ {\rm 0.1 m}} \right), \nn
\label{eq:ba}
\eea
where $R_1$ is the inner radius of the capacitor and $E_0$ denotes the strongest electric field strength at the inner radius. $c_R$ is an ${\cal O}(1)$ parameter determined by the geometry of the capacitor cross-section. $v_{\rm DM}=10^{-4}$ is the the Earth's local relative velocity to the dark matter halo. For instance, for a two-layer cylindrical capacitor with inner and outer radii as $R_1$ and $R_2$, $c_R = \ln (R_2/R_1)$.

An $N_1$-turn coil of radius $R_{\rm coil}\sim R_1$ is placed in the central region of the capacitor to pick up the induced magnetic flux $\Phi_a=B_a\cdot \pi R_1^2$. This magnetic flux corresponds to the $B_a$ component parallel to the cylinder axis, and is proportional to the $v_{a,z}$ component along the $\hat{z}$; the $\vec{v}_a$ projections in the $\hat{x}$ and $\hat{y}$ direction do not contribute to $\Phi_a$. In our calculations we would assume $\vec{v}_a //{\hat{z}}$, while in measurement $\vec{v}_a\cdot{\hat{z}}$ yields a directional dependence in $\Phi_a$ that in principle can be used as check by adjusting the experiment's orientation.

The pickup coil is connected as part of a LCR circuit with it resonance point tuned to a target search frequency of the axion field oscillation, e.g. $LC =\omega^{-2}$. The total inductance $L$ (without subscript) includes contributions from the pickup coil $L_1$, the effective inductance $L_2$ of the inductive coupling to amplifier circuit and any additional inductance from wiring. $L_2$ includes both the coupled coils' self-inductance and mutual inductance. We would require $L_1 \gg L_2$ to ensure $L\sim L_1$. Note a single-turn circular loop with wire radius $d_1\sim $mm has self-inductance around $10 \mu$H per meter. The pickup coil $L_1$ can be loosed winded to reduce the inductance between its adjacent turns, so that $L_1\propto N_1$. The LCR resonance requires a capacitance value $C=(2\pi f)^{-2}/L\sim 0.3{\rm pF}\cdot(\mu{\rm H}/L)\left({\rm 0.1 ~GHz}/{f}\right)^2$, which increases toward lower axion mass, and is much larger than typical stray capacitance on meter-scale wiring. Low-temperature tuning of the LCR resonance frequency can be achieved via an adjustable capacitor.

The high LCR quality factor at its resonance point allows for an enhanced current in the circuit~\cite{Sikivie:2013laa}. The saturated current in the  loop is
\be
I_a =Q_c \cdot  (\pi R_1^2 N_1 B_a L^{-1}) \cos \omega t,
\ee
where $Q_c$ denote the conventional LCR quality factor $Q_c=\omega_a L/R_s$. The signal power is stored in the LCR circuit, till the energy dissipation power by the resistance is $P_{\rm dis.}= Q_c\cdot(N_1\Phi_a/L)^2\omega_a L/2$ equals that of the signal feed. For efficient signal pickup, $Q_c$ should match to the quality factor in the axion field $Q_c\sim Q_a$, which also matches the saturated energy dissipation to the dark matter axion-photon energy conversion rate. Note for a high $Q_c$, saturating the maximal current takes $Q_c$ oscillation periods and it could require significantly longer time $Q_c\cdot 2\pi\omega_a^{-1}$ in case of extremely low axion masses. In the MHz$\sim $GHz range, saturation time is less than a second and we consider the maximal LCR current is always reached during the observation time.

The inner volume of the cylindrical capacitor, along with the pickup circuit, is placed under cryogenic conditions to suppress the thermal noise. Environmental fluctuations can be removed via electromagnetic shielding. Since there is no strong magnetic field in the system, the inductance coils and the capacitor in the LCR loop can be made with superconductor materials, such that the circuit components with the major resistance can be confined to a small spatial region and placed under an extreme cryogenic temperature, $T_c\sim$ mK. This reduces the cooling requirement on the sizable superconducting pickup $L_1$, without worthening its noises. For instance, NbTi superconductor has a transition temperature at 9.7 K and has good transmission rate for $10^2$ MHz signals. Denoting $R_s$ as total resistance in the LCR circuit, a low effective resistance on the superconducting NbTi pickup coil $R_{\rm coil}\ll R_s$ would allow a much higher temperature $T_{\rm coil} = T_c\cdot(R_s/R_{\rm coil})$ and keep the coil's thermal noise to be subdominant. Note for a meter-radius, loosely winded 10 round loop, $L_1\sim 50-100 \mu$F and the LCR resonance requires $R_s=\omega L/Q_c=~0.04{\rm \Omega}\cdot (f/{\rm GHz})$. Contact resistance can reduced to $10^{-3}~{\rm \Omega}$. If $L_1$ is at $1$ K with the rest of the LCR at mK, $L_1$'s effective resistance needs to be controlled to $R_{\rm coil}<10^{-5}~{\rm \Omega}$.

\section{Noise \& Sensitivity}
\label{sect:sensitivity}

The induced current $I_a(t)$ in the LCR circuit is then coupled to the amplifier/detector chain. The leading current noise in the LCR is the Johnson-Nyquist (JN) noise due to the resistance in the LCR circuit, $\delta I^2= k_B T_c \Delta f/R_s$.  This noise will also be amplified at the subsequent amplifier(s) and can be the major noise in signal measurement. The total noise power is
\be
P_n = k_B T_c \Delta f +  k_B T_D \Delta f,
\ee
where the latter is an equivalent readout noise power that includes the added noises from amplifier and detector chain, etc. For the detection bandwidth we take the optimal choice $\Delta f = Q_c^{-1} f$. 
High sensitivity, low dissipation amplication helps maintain the SNR from the LCR loop. For instance, using a SQUID to convert the inductively coupled $I_a(t)$ into to voltage signal is proposed in Ref.~\cite{Sikivie:2013laa}, and other amplifications are possible. State-of-the-art cryogenic detectors like the HEMT have $T_D\sim K$ noise temperature. Recent developments in interferometry techniques can further reduce the detector noise level to $T_D \le 0.1$ K~\cite{Bozyigit2010,Peng2016} and is applicable in axion search~\cite{Yang:2022uil,Mcallister:2019jgj}. With sufficient amplification it is possible for the amplified JN noise to be larger than the added detector noise. Thus for a proof-of-principle analysis in this paper, we take $T_c =1$~mK and $10~$ mK as benchmark scenarios and assume the pickup circuit's thermal noise dominates after amplification.

Assuming thermal noise dominates after amplification, the SNR for the signal with $I_a^2$ and $\delta I^2$ is
\bea
{\rm SNR} &=& \frac{(Q_c N_1 \Phi_a/L)^2R_s}{2 k_B T_{c}}\sqrt{\frac{t}{\Delta f}}~~,\nn \\
 &=& \frac{Q_c ( N_1\cdot \pi R_1^2B_a)^2}{2 Lk_B T_{c}}\sqrt{Q_c \cdot 2\pi \omega_a\cdot t}
 \label{eq:snr}
\eea
where $t$ is observation time, $g$ is the effective signal power gain from subsequent amplifiers, and we used the relation $R_s=L\omega_a/Q_c$ at the resonance frequency. This SNR formula is similar to that of a resonant cavity haloscope in the sense of a high $Q$ enhancement from front-end frequency filtering on the near-monochromatic signal.

\begin{figure}[t]
\includegraphics[scale=0.59]{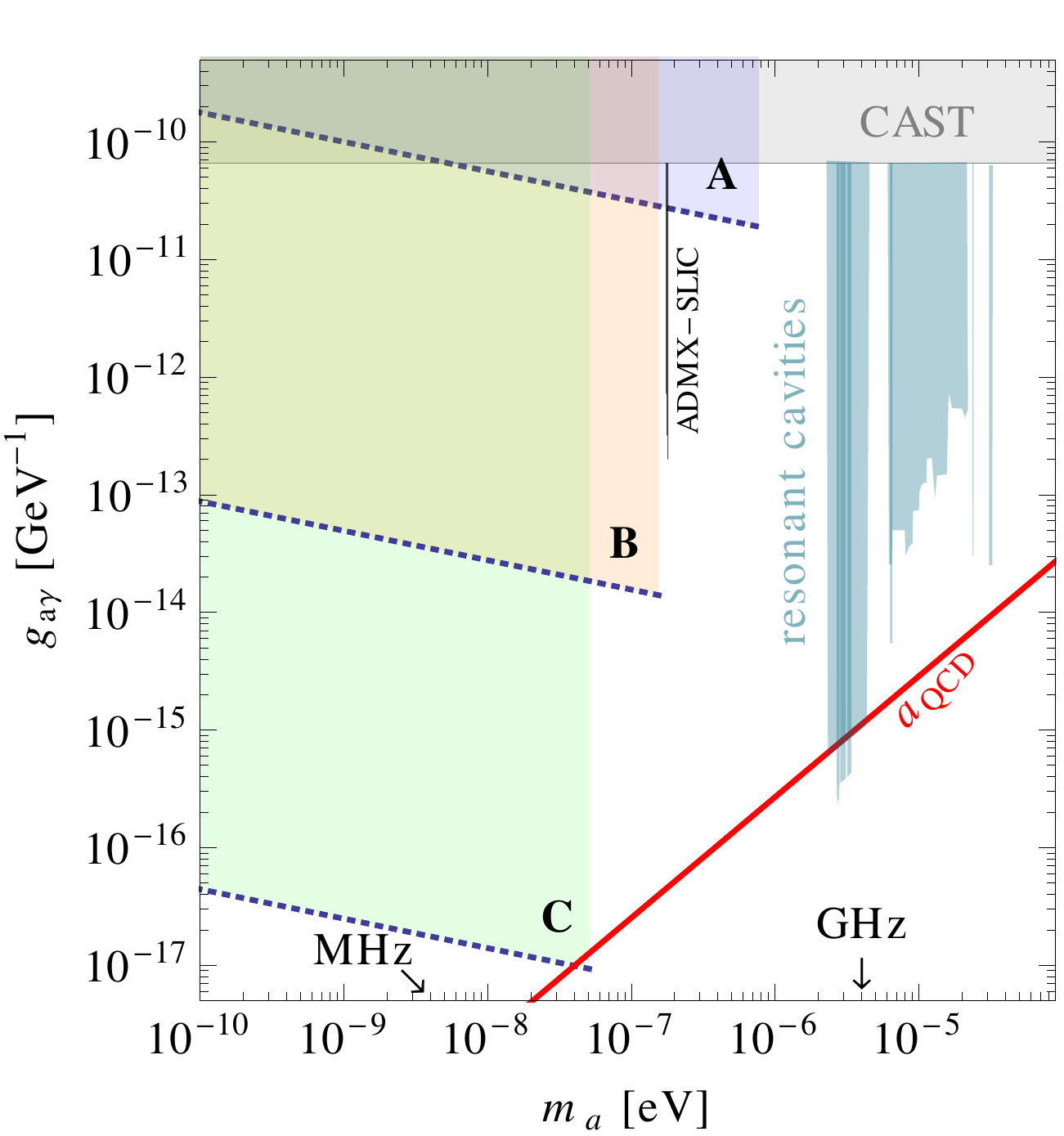}
\caption{Prospective sensitivity with benchmark parameter sets listed in Table~\ref{tab:benchmarks}. The projected $g_{a\gamma}$ sensitivities (dotted lines) assumes one week observation time at each frequency, and the maximal $m_a$ corresponds to the axion field coherence limit that $\lambda_\gamma > 8R_1$. The CAST limit (gray, top) and recent results from cavity haloscopes (vertical, right), ADMX-SLIC (vertical, middle) are shown for comparison.}
\label{fig:reach}
\end{figure}

In Eq.~\ref{eq:snr}, lowering $L$ helps but it is related to the pickup loop's own geometry. When the $L_1$ coil is loosely winded, $L\approx N_1\times R_1\left[\ln{(8R_1/r_1)}-2\right]$ and the SNR still benefits from increasing the turn number $N_1$ if $L_1$ dominates the total $L$. We take modest numbers $N_1=5,10$ as it is desirable to have a large $R_1$ to maximize the pickup area.
The benchmark selections of experimental parameters are given in Table~\ref{tab:benchmarks}. Demanding SNR=3 and observation time $t$ on one frequency point, the sensitivity on the axion-photon coupling is
\bea
g_{a\gamma}&=&\frac{\sqrt{{\rm SNR}\cdot 2N_1L_{1,0}\cdot k_B T_c}}{(\pi R_1^3 N_1 E_0 v_a c_R\sqrt{2\rho_{\rm DM}}) \sqrt[4]{Q_c^3~ 2\pi \omega_a t}} \nn ,\\
 &\approx & 1.6\times 10^{-12}{\rm ~GeV^{-1}} \left(\frac{R_1}{1 ~{\rm m}}\right)^{-3}  \left(\frac{E_0}{\rm MVm^{-1}} \right) \nn \\
 &\times & \left(\frac{m_a}{10^{-6}~{\rm eV}}\cdot \frac{t}{\rm hr} \right)^{-1/4},
 \label{eq:reach}
\eea
where $N_1=5$ and $L_{1,0}\equiv L_1/N_1\sim \mu$H represents the inductance per turn. The sensitivity reach with benchmark parameters are illustrated in Fig.~\ref{fig:reach}. Benchmark A assumes modest parameter values that can fit into conventional dilution fridges, and for $t\sim 1$ week observation time, the sensitivity dips below the current solar axion search constraint~\cite{Anastassopoulos:2017ftl} at the optimal frequency around 0.2 GHz. Benchmark B assumes more aggressive cryogenic conditions for a multi-meter cylinder and the projected sensitivity can reach $g_{a\gamma}\sim 3\times 10^{-15}$~GeV$^{-1}$ at $m_a=0.1~\mu$eV, or $19$ MHz. Benchmark C shows that an electric field strength around GVm$^{-1}$ is required in order to touch down to the theoretical prediction for the (KSVZ) QCD axion, which can be achieved by using high $E$-field tolerant material that is transparent in the relevant waveband, such as diamond, etc. As shown in Eq.~\ref{eq:reach}, the sensitivity improves towards larger $m_a$, but it has its coherence limit: the diameter of the electrified cylinder should not exceed the half-wavelength of the induced electromagnetic wave $\lambda_\gamma =m_a^{-1}/2$, or $8R_1> \lambda_a$.

\begin{table}[h]
\begin{tabular}{c|cccc}
\hline
Benchmark & ~$R_1$(m)~ & ~$N_1$~ & ~$E$(V/m)~ & ~$T_c$(mK)\\
\hline
A & 0.2 & 5  &  $10^6$  & 10 \\
B &   1 & 10 &  $10^7$  & 1  \\
C &   3 & 20 &  $10^9$  & 1  \\
\hline
\end{tabular}
\caption{Benchmark setups. Common parameters include $Q_c\sim 10^6$, $\rho_{\rm DM}=0.4$ GeV cm$^{-3}$, pickup wire radius $r_1=1$ cm. Corresponding $g_{a\gamma}$ sensitivity is shown in ~Fig.~\ref{fig:reach}.}
\label{tab:benchmarks}
\end{table}

Apparent from Eq.~\ref{eq:reach}, increasing the cylinder size and/or the electric field strength are efficient for the purpose of improving the $g_{a\gamma}$ sensitivity. This corresponds to a stronger induced $\vec{B}_a$ field. Improvements from more pickup coil turns, lowering the noise temperature and a longer observation time are less effective. The $g_{a\gamma}$ limit scales with the quality factor as $Q^{-3/4}$, which shows the necessity of exploiting the dark matter's coherence via proper frequency filtering.

Although axion dark matter converts faster in a strong magnetic field, as at $v_{\rm DM}=10^{-4}$, $B=$10 T is equivalent to $E\sim 3\times 10^{4}$ GVm$^{-1}$ when it comes to the $a\rightarrow\gamma$ conversion rate. However, one major advantage of using the electric field as the axion-photon conversion medium is to avoid exposing the sensitive pickup to an applied magnetic field. For example, the magnetic fluctuations would become an intrinsic noise that requires careful treatment~\cite{ADMXSLIC}. For the cylindrical capacitor, high-resistance material can be used to reduce the current noises on its surfaces, and electric leakage is a DC current that does not leave a lasting remnant in the filtered frequency range.

\section{Summary \& Discussion}
\label{sect:discussion}

In this paper we explore the concept of a narrow-band axionic dark matter detection scheme with a strong laboratory static electric field as the $a\rightarrow \gamma$ converting medium. The geometry of a cylindrical capacitor allows the axion-induced effective current to form a solenoid pattern and yield a concentrated magnetic field signal in the inner region of the capacitor. As axion-photon conversion is suppressed by axion velocity under static electric field, we adopt a narrow band approach to enhance the signal detection. Working at relatively long wavelengths, an LC resonance offers the front-end high $Q$-factor frequency filtering, which can be matched to the dark matter field's energy dispersion. The high $Q$-factor allows the signal to build up in the LCR circuit that effectively enhances the signal strength, and can be probed by cryogenic amplifiers and detectors with high sensitivity. 

We propose three benchmark scenarios, and demonstrate that a sub-meter capacitor scheme with mainstream cryogenic technology is sensitive to $g_{a\gamma}\sim 10^{-12}$ GeV$^{-1}$. For a multi-meter cylindrical capacitor and mK-level noise temperature, the projected sensitivity can be further improved to $10^{-15}$ GeV$^{-1}$ scale with $10 $~MVm$^{-1}$ static $E$ field strength. The theoretical QCD axion prediction is achievable with a high field $E\sim 10^9$ Vm$^{-1}$ setup that involves further R\&D on implementing high dielectric materials.

Besides avoiding magnetic fluctuations, the absence of a strong magnetic field allows the pickup coils to use superconductor materials that reduces the LCR circuit's resistance. For thermal noise control, only the circuit parts with resistance needs to be restricted to the lowest ($\sim$mK) temperature. Given a meter-scale size of the cylinder, this reduces the overall cooling requirement of the system.

Due to the direction dependence in the effective $\vec{j}_a=-g_{a\gamma}a\vec E\times \vec{k}_a$, the signal strength with static electric field is modulated by both axion field periodicity and apparatus orientation. Adjustment of capacitor direction, as well as daily modulation and be used as an additional method of differentiation of the axion-induced signal from the system's own systematics.

\medskip
{\bf Acknowledgment.}
~Authors thank Zhihui Peng and Yi Zhou for helpful discussions.
Y.G. is supported under grant No. 12150010 supported by the National Natural Science Foundation of China, and in part by the Institute of High Energy Physics, CAS (E2545AU210), and the Ministry of Science and Technology of China (2020YFC2201601). Q.Y. is supported by the National Natural Science Foundation of China under Grant No. 11875148, and No. 12150010.

\bibliography{refs}

\end{document}